\documentclass{article}

\usepackage{PRIMEarxiv}

\usepackage[utf8]{inputenc} 
\usepackage[T1]{fontenc}    
\usepackage{hyperref}       
\usepackage{url}            
\usepackage{color}
\usepackage{booktabs}       
\usepackage{amsfonts}       
\usepackage{nicefrac}       
\usepackage{microtype}      
\usepackage{lipsum}
\usepackage{fancyhdr}       
\usepackage{fancyvrb}
\usepackage{graphicx}       
\usepackage{amsmath}
\usepackage{mhchem}
\usepackage{makecell}
\usepackage[ruled,vlined,linesnumbered]{algorithm2e}
\usepackage{upgreek}
\usepackage[backend=biber,style=nature,maxnames=3,minnames=1,sorting=none]{biblatex}
\newcommand{\micron}{$\upmu$m}

\newcommand{\norm}[1]{\left\lVert#1\right\rVert}
\newcommand{\admmpenalty}{\tau}

\DeclareMathOperator*{\argmin}{arg\,min}

\title{Pty-Chi: A PyTorch-based modern ptychographic data analysis package}

\author{
  Ming Du$^*$, Hanna Ruth, Steven Henke, Yi Jiang, Viktor Nikitin, Ashish Tripathi, \\
  \textbf{Junjing Deng, Jeffrey Klug, Peco Myint} \\
  Advanced Photon Source \\
  Argonne National Laboratory \\
  Lemont, IL, USA \\
  $^*$ \texttt{mingdu@anl.gov} \\
\AND
  Tao Zhou \\
  Center for Nanoscale Materials \\
  Argonne National Laboratory \\
  Lemont, IL, USA \\
\AND
  Nicholas Schwarz, Mathew Cherukara$^\dagger$, Alec Sandy, and Stefan Vogt \\
  Advanced Photon Source \\
  Argonne National Laboratory \\
  Lemont, IL, USA \\
  $^\dagger$ \texttt{mcherukara@anl.gov}
}

\begin{document}

\textbf{GOVERNMENT LICENSE}

The submitted manuscript has been created by UChicago Argonne, LLC, Operator of Argonne
National Laboratory (``Argonne''). Argonne, a U.S. Department of Energy Office of Science laboratory, is operated under Contract No. DE-AC02-06CH11357. The U.S. Government retains for
itself, and others acting on its behalf, a paid-up nonexclusive, irrevocable worldwide license in
said article to reproduce, prepare derivative works, distribute copies to the public, and perform
publicly and display publicly, by or on behalf of the Government. The Department of Energy will
provide public access to these results of federally sponsored research in accordance with the DOE
Public Access Plan. http://energy.gov/downloads/doe-public-access-plan.

\maketitle

\begin{abstract}
Ptychography has become an indispensable tool for high-resolution, non-destructive imaging using coherent light sources. The processing of ptychographic data critically depends on robust, efficient, and flexible computational reconstruction software. We introduce Pty-Chi, an open-source ptychographic reconstruction package built on PyTorch that unifies state-of-the-art analytical algorithms with automatic differentiation methods. Pty-Chi provides a comprehensive suite of reconstruction algorithms while supporting advanced experimental parameter corrections such as orthogonal probe relaxation and multislice modeling. Leveraging PyTorch as the computational backend ensures vendor-agnostic GPU acceleration, multi-device parallelization, and seamless access to modern optimizers. An object-oriented, modular design makes Pty-Chi highly extendable, enabling researchers to prototype new imaging models, integrate machine learning approaches, or build entirely new workflows on top of its core components. We demonstrate Pty-Chi’s capabilities through challenging case studies that involve limited coherence, low overlap, and unstable illumination during scanning, which highlight its accuracy, versatility, and extensibility. With community-driven development and open contribution, Pty-Chi offers a modern, maintainable platform for advancing computational ptychography and for enabling innovative imaging algorithms at synchrotron facilities and beyond.
\end{abstract}


\section{Introduction}

Coherent diffraction imaging (CDI) techniques \cite{Hawkes2019-wa, Jacobsen2019-iq} rely on phase retrieval computations to yield human-understandable images from measured diffraction patterns. As such, a performant, robust, and efficient reconstruction package is as crucial to the final quality of the images as the measured data. 

Ptychography is a scanning variant of CDI \cite{Rodenburg2004-nw} that removes the limit on the field of view size in traditional full-field CDI techniques. This characteristic leads to its widespread application with coherent light sources for high-resolution, non-intrusive imaging of extended samples. Various ptychography packages exist which feature one or multiple analytical reconstruction algorithms with hand-crafted update rules and preconditioning such as the ptychographic iterative engine (PIE) family \cite{Rodenburg2004-nw, Maiden2009-md, Maiden2017-um}, least-square maximum likelihood (LSQML) \cite{Odstrcil2018-ns}, difference map \cite{Elser2003-ov} (DM), conjugate gradient descent (CG) \cite{Tripathi2014-nv}, and the Bilinear Hessian (BH) method \cite{Carlsson2025-xw}. Examples of these packages include PtychoShelves \cite{Wakonig2020-ap}, PtychoLib \cite{Nashed2014-eh}, PtychoPy \cite{Yue2021-ag}, PtyPy \cite{Enders2016-vs}, and PyNX \cite{Favre-Nicolin2020-cj}. Recently, deep learning packages such as PyTorch \cite{Paszke2019-xm} and TensorFlow \cite{Abadi2016-rf} have accelerated the adoption of automatic differentiation (AD) in computational phase retrieval, fostering packages such as Adorym \cite{Du2021-ra} and DAP \cite{Wu2024-sg}. Gradient-based phase retrieval built with AD leaves the derivation of the gradients of the unknowns to the AD engine, so that users can flexibly make changes in the forward models without re-deriving the gradients. The use of modern deep learning libraries also allows developers to access a collection of built-in optimizers and easily swap them out for better performance. 

Each ptychographic reconstruction algorithm comes with its own benefits and limitations, and therefore a full-fledged ptychographic reconstruction package should provide a comprehensive set of choices for users to adapt to their own scenarios. AD-based methods are flexible, extendable with a wide selection of optimizers, and ideal for prototyping methods to solve the reconstruction problem with new unknowns in the image formation model, but analytical algorithms that are well-crafted for ptychography still outperform AD methods on some occasions. For example, the ``compact'' batch partitioning in LSQML \cite{Odstrcil2018-ns} delivers better accuracy and is more resistant to grid pathology \cite{Thibault2009-bs} than AD particularly for images that contain large areas of empty space. Also, the object preconditioning in LSQML formed by superimposing the illumination function in the object frame introduces the overlapping constraint unique in ptychography, leading to better performance for undersampled data \cite{Odstrcil2018-ns}. One analytical algorithm also possesses unique merits compared to another. For example, alternating projection algorithms like DM often demonstrate faster convergence but use more memory. Algorithms in the PIE family are relatively simple and demand fewer computing resources. 

Runtime speed is another important criterion of ptychography packages in view of the ever-increasing data size. Being able to utilize hardware accelerators (often GPUs) is essential for modern ptychographic data processing. In addition, being agnostic about accelerator vendors is another major bonus as it allows users to launch reconstruction jobs with devices made by different manufacturers such as NVIDIA, AMD, and Intel. Packages that build most of their computation routines with a specific programming model like CUDA can be very performant on devices of compatible vendors, but they lose the adaptability to other devices. Moreover, multi-device parallelization is also a desirable feature as it speeds up large-scale computation and eases the memory pressure in reconstructing large datasets. 

Maintainability and accessibility are equally crucial criteria for the adoption of a ptychography package in the open science community. Packages developed in proprietary programming languages are unfortunately faced with significant barriers in their use as the interpreters of the languages often require an expensive paid license to run. It is also because of these barriers that experienced developers able to contribute to and maintain the packages are often in short supply. On the other hand, Python has been democratizing high-performance scientific programming for years. An extremely large and fast-growing community is already present, and a sophisticated software ecosystem, including PyTorch and TensorFlow mentioned above, has been established. A ptychography package built on Python is undoubtedly easier to be widely adopted and improved on by the community.  

Computational imaging is a fast evolving domain especially with the adoption of artificial intelligence (AI) techniques. As such, we also emphasize the extendability of reconstruction packages. Packages with a modular design allow one to subclass a certain component (such as the reconstructor, forward model, and the data structures containing the object, probe, and other parameters) to implement a novel variant, then plug it into the package. This also allows one to develop other algorithms and packages with totally different workflows using the data structures and/or components of the said package, as the object-oriented design makes the reconstruction parameters and internal states accessible, allowing them to be modified externally after every epoch of reconstruction.

We present Pty-Chi, a modern ptychographic reconstruction package that features the following characteristics:
\begin{itemize}
    \item Pty-Chi is built on PyTorch \cite{Paszke2019-xm}, a community-maintained device-accelerated deep learning framework that has a very large user base. Pty-Chi's computation routines are written with PyTorch's mathematical functions; variables are stored as PyTorch's tensor data structure, and modules (such as the forward model) inherit PyTorch's \texttt{nn.module} class. The direct benefits are two-fold: first, many operations and components are differentiable, allowing an AD reconstructor to be created; the differentiable forward model of Pty-Chi also allows users to develop machine learning algorithms based on it. Second, we automatically get the hardware accelerator support of PyTorch. With Pty-Chi itself being vendor-agnostic, one can adapt to devices from various vendors by simply installing PyTorch built against the libraries for those vendors.

    \item Pty-Chi features classical and high-performance reconstruction algorithms including ePIE \cite{Maiden2009-md}, rPIE \cite{Maiden2017-um}, DM \cite{Elser2003-ov}, LSQML \cite{Odstrcil2018-ns}, and the Bilinear Hessian method (BH) \cite{Carlsson2025-xw, Carlsson2025-vr}. It also supports features that account for non-ideal experiment conditions, which include probe position correction, orthogonal probe relaxation (OPR) \cite{Odstrcil2016-wy}, and multislice modeling \cite{Maiden2009-md}. Additional techniques that have been proven effective in improving reconstruction quality, such as the compact batching method introduced in \cite{Odstrcil2018-ns}, are also available. These features are mostly orthogonal to the reconstruction algorithms, meaning they are largely reconstructor-agnostic and applicable to most reconstructors if mathematically compatible.

    \item Pty-Chi also allows the use of most optimizers included in PyTorch, such as Adam \cite{Kingma2014-dm} and its variants, for the update of reconstruction parameters. These optimizers are also implemented orthogonal to the reconstructors instead of bound to any of them, which makes them generally applicable across different reconstruction algorithms. 

    \item Pty-Chi employs an object-oriented architecture. In addition to better customizability, modularity and ease of maintenance, a major advantage is that since the reconstruction task manager and reconstructor are objects with accessible internal states, users may easily read and modify these internal states between reconstruction epochs without changing Pty-Chi itself to realize extended capabilities. Some examples will be shown later in this paper. 

    \item Pty-Chi also provides multi-GPU and multi-node parallelization for selected reconstructors. This is realized through multi-processing utilities in \texttt{torch.distributed} (for analytical reconstructors) and the \texttt{DistributedDataParallel} wrapper (for AD).

    \item Finally, Pty-Chi also provides extended capabilities involving non-traditional reconstruction approaches. For example, it allows the object function to be represented by a noise-to-image neural network; combined with the differentiable forward model and AD reconstructor, this achieves deep image prior-based reconstruction \cite{Lempitsky2018-ee} which often performs better with low probe overlap. We will also show an example where a different package is developed using Pty-Chi as a dependency, where it inherits the data structures of Pty-Chi, runs Pty-Chi's reconstructor, and gets and modifies the reconstructed object after every set number of epochs. 
\end{itemize}

We show the design, results of challenging test cases, and creative usage of Pty-Chi. We have published the source code of Pty-Chi on GitHub (\url{https://github.com/AdvancedPhotonSource/Pty-Chi}) where a detailed documentation site is linked.

\section{Software design}

\subsection{Overview}

\begin{figure}
    \centering
    \includegraphics[width=1\linewidth]{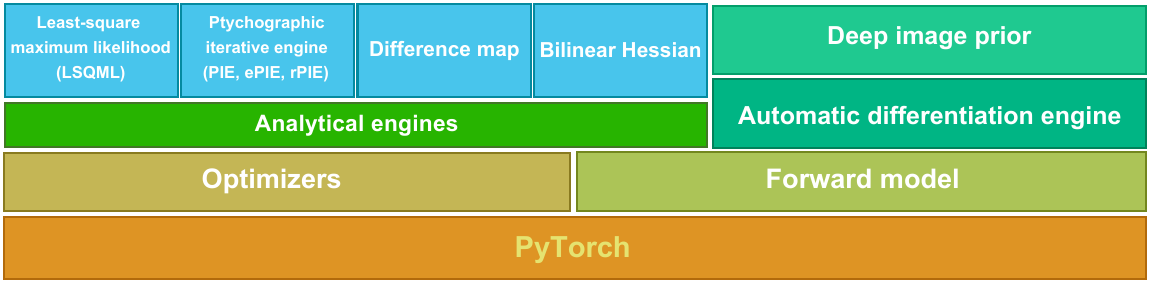}
    \caption{Main building blocks and layers of Pty-Chi.}
    \label{fig:hierarchy}
\end{figure}

We provide a high-level overview of the most essential building blocks of Pty-Chi in Fig.~\ref{fig:hierarchy}. PyTorch is the backend engine for basic mathematical routines, data structures, and GPU acceleration. Based on these, we built a ptychographic forward model shared by all reconstructors, which
\begin{itemize}
    \item is end-to-end differentiable,
    \item handles multiple scan points and probe modes in a vectorized fashion,
    \item supports scan point-specific probes (needed by OPR), multislice modeling, and both far-field and near-field for free space propagation.,
    \item is a sub-class of \texttt{torch.nn.Module}, allowing the forward model object and all its registered buffers to be moved across GPUs as a whole, and allowing it to be wrapped by \texttt{DataParallel} or \texttt{DistributedDataParallel} to enable multi-GPU or multi-node parallelization.
\end{itemize}

Meanwhile, we allow users to select from the built-in optimizers of PyTorch for parameter updates. The optimizers, along with the forward model, support both analytical and AD reconstructors. For analytical reconstructors, the default optimizer is steepest gradient descent with a step size of 1, so that it does not change the designed behavior of these algorithms. More sophisticated optimizers can be used in conjunction with the update vectors calculated by analytical algorithms. For example, using a momentum-based optimizer for object and probe update in rPIE essentially transforms it to mPIE \cite{Maiden2017-um}. These advanced optimizers can also be applied to other reconstruction parameters, such as probe positions, which have been shown to lead to faster convergence \cite{Liu2025-dynamic}.

\begin{figure}
    \centering
    \includegraphics[width=1\linewidth]{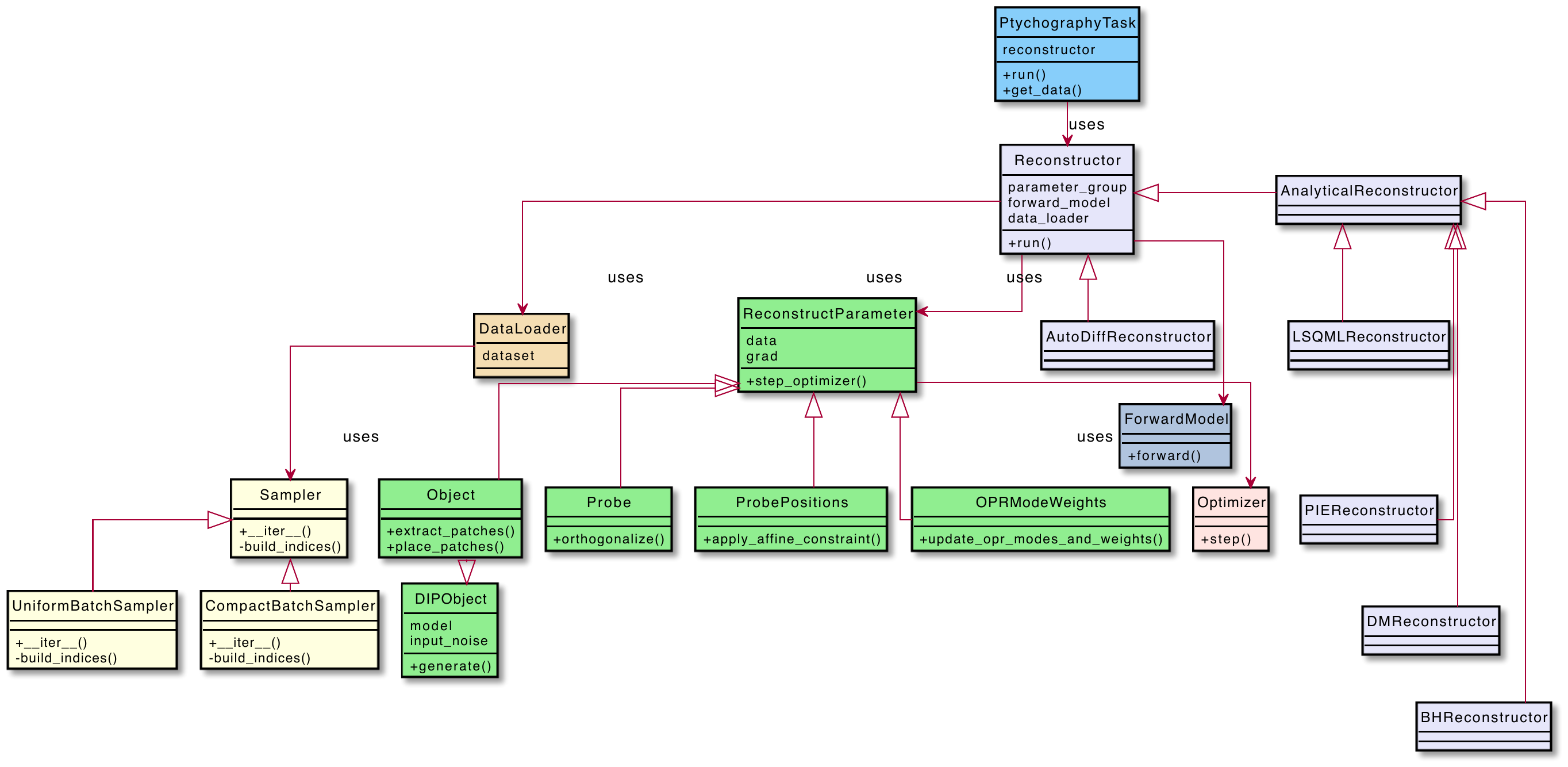}
    \caption{The UML diagram of Pty-Chi.}
    \label{fig:uml}
\end{figure}

We show the unified modeling language (UML) diagram of Pty-Chi in Fig.~\ref{fig:uml}, which covers a broader range of components. Reconstruction parameters, such as object, probe, positions, and OPR mode weights, have distinct classes. Routines shared across all reconstructors, including object patch extraction/update and probe mode orthogonalization, are implemented as methods of these classes. The object class is subclassed by a \texttt{DIPObject} class that implements a neural network-parameterized variant of the object function, allowing for deep image prior-based reconstruction \cite{Lempitsky2018-ee} with the AD reconstructor.

Another component implemented orthogonally to the reconstructors is the batch sampler. A sampler generates the indices of scan points to be processed in each minibatch. In addition to simple random sampling, we also provide the uniform sampler that returns a batch of scan points that have a uniform density all over the space, and the compact sampler that returns a spatially connected cluster of scan points each time. Both techniques follow the description in \cite{Odstrcil2018-ns}. The samplers are used by the native \texttt{DataLoader} of PyTorch, allowing us to use its available features such as memory pinning. 

\subsection{Reconstructors}

We first visit the AD reconstructor of Pty-Chi. With a differentiable forward model predicting the detector-plane intensity and a loss function computing the discrepancy between the predicted and measured intensities, the AD engine PyTorch calculates the gradient of the loss with regards to every optimizable variable, storing the gradient in their tensors; the optimizer then updates the variables using the gradients. AD automatically responds to changes in the forward model or the parameters being reconstructed, which makes it ideal for prototyping a new algorithm that models and corrects additional ``moving parts'' in the image formation process. For example, by making the slice spacing of a multislice object optimizable, we have conveniently realized slice spacing correction with the AD reconstructor. 

However, the gradients directly coming from AD often contain data-specific scaling, making it necessary to tune the step size over a large range every time. For instance, consider a mean squared error (MSE) loss on magnitudes:
\begin{equation}
    L = \frac{1}{B\times H\times W}\sum_{b=1}^B \sum_{j=1}^{H\times W}\left(\sqrt{I_\mathrm{pred}} - \sqrt{I_\mathrm{meas}}\right)^2
\end{equation}
where $B$ is the batch size, $H$ and $W$ are the height and width of a diffraction pattern. The direct gradient of $L$ with respect to the patches of the object function is
\begin{equation}
    \frac{\partial L}{\partial o} = -\frac{1}{B\times H\times W} \left[ p^* (\psi' - \psi) \right]
\end{equation}
where $p^*$ is the complex conjugate of the probe, $\psi'$ is the magnitude-updated wavefield back-propagated to the exit plane, and $\psi = p\cdot o$ is the exit wave during forward propagation. This gradient has two issues: first, the prefactor of $1/(B\times H\times W)$ is dependent on the batch size and the size of the diffraction pattern; second, the gradient's magnitude also depends on the scale of $p$, which often varies. To make the scale of gradients more consistent, we adopt the treatment in ePIE and scale the object gradient by $1/\mbox{max}|p|^2$; we also multiply the gradient by $B\times H\times W$ to undo the influence of the pre-factor. Similar treatments are also applied to other variables to ensure the consistency of the scales of their gradients.

The analytical reconstructors of Pty-Chi current include the PIE family \cite{Rodenburg2004-nw, Maiden2009-md, Maiden2017-um}, LSQML \cite{Odstrcil2018-ns}, DM \cite{Elser2003-ov}, and BH \cite{Carlsson2025-xw}. The convergence behaviors of these algorithms have been compared in some of these cited works. In practice, the choice of reconstructors to use is typically made based on the following factors:
\begin{itemize}
    \item Reconstructors based on alternating projection (DM) and those with self-adaptive step sizes (LSQML, BH) are easy to transfer across datasets without additional parameter tuning. 
    \item Minibatched (another notion for ``ordered subset'' \cite{Hudson1994-ky}) reconstructors are lightweight and less memory consuming, though alternating projection algorithms like DM can also reduce peak memory usage by accumulating batch-wise gradients before each update.
    \item Some reconstructors take special treatment given a certain batch sampler to further improve the result. For example, following the implementation in PtychoShelves \cite{Wakonig2020-ap}, LSQML accumulates the update vector of the object and only updates the object once per epoch when the compact batch sampler is used, which empirically leads to better contrast for objects containing large areas of empty space. Also, for compact batching, the LSQML reconstructor can apply a special momentum acceleration for object and probe updates that are specifically crafted for this mode (different from the classical momentum-accelerated gradient descent optimizers provided by PyTorch. We plan to generalize and encapsulate this momentum acceleration method into a subclass of \texttt{torch.optim.Optimizer} so that it can be easily used by other reconstructors).
\end{itemize}

\subsection{Parallelization}

While all reconstructors support GPU acceleration, multi-GPU support is currently only available with the AD reconstructor and a subset of analytical reconstructors. We plan to extend parallel computing support to more analytical reconstructors upon high user demand. 

Pty-Chi adopts data parallelism, where a copy of reconstruction parameters such as the object and probe is stored on each GPU. The scan points in a minibatch are split evenly and sent to each GPU. All GPUs asynchronously compute the gradients or update vectors using the fraction of data they receive. Before each update, the gradients are collected from all GPUs and summed. Data parallelism is clean to implement, easily maintainable, and introduces few alterations to single-GPU code. Although it results in redundant memory usage for storing copies of reconstruction parameters on every GPU, this additional memory consumption is only fractional compared to the memory required by the vectorized processing of the entire minibatch of data involving multiple probe modes. We also note that although each GPU calculates the gradients using only $B / N_{\textrm{devices}}$ diffraction patterns at a time, the gradients from all GPUs are summed (synchronized) before the parameters are updated, so this is mathematically equivalent to updating the gradient with $B$ diffraction patterns.

\subsection{Extension through creative usage}

Rather than providing the reconstruction routine as a black box with only an entry point to users, Pty-Chi's object-oriented design ensures the reconstruction task manager object and all the components in it (Fig.~\ref{fig:uml}) are accessible and modifiable. This allows users to build additional capabilities on the existing code of Pty-Chi through creative workflow design. The following pseudocode shows an example of how one can conveniently realize the reconstruction of a single object function using 2 sets of diffraction patterns, each with their own distinct probe. A task manager is created for each dataset, which has its own probe, positions, and OPR modes; at each outer epoch, each task manager is run for only one epoch. When the epoch finishes, all internal states and parameters of the task are preserved. One then copies the updated object function to the next task manager (if the current one is the last, copy it to the first one). The object is thus sequentially updated by each task manager with its own data and probe, and the loop continues until the set number of outer epochs is reached.
\begin{Verbatim}[frame=single]
# Create options and task manager object for dataset 1
options_1 = LSQMLOptions()
...
task_1 = PtychographyTask(options_1)

# Create options and task manager object for dataset 2
options_2 = LSQMLOptions()
...
task_2 = PtychographyTask(options_2)

all_tasks = [task_1, task_2]
for _ in range(options_1.reconstructor_options.num_epochs):
    for i_task, task in enumerate(all_tasks):
        # Run one epoch with the current task manager
        task.run(1)
        
        # Copy updated object to the next task circularly
        i_next_task = (i_task + 1) % len(all_tasks)
        all_tasks[i_next_task].copy_data_from_task(task, params_to_copy=("object",))
\end{Verbatim}

The following example reveals a potential way of realizing joint ptycho-tomography reconstruction using Pty-Chi, where 3D reconstruction capability is not inherently provided, following the strategy proposed in \cite{Gursoy2017-hk}. No changes inside Pty-Chi are needed; users only need to implement their own functions for 3D projection and backprojection. 
\begin{Verbatim}[frame=single]
# We assume a 2D ptychographic dataset is collected at each viewing
# angle, and create the options and task manager for each of them.
all_tasks = []
for i_angle in range(n_angles):
    options = LSQMLOptions()
    ...
    task = PtychographyTask(options)
    all_tasks = []

# Create a 3D object buffer.
object_3d = torch.ones([n_x, n_y, n_z]).complex()

for _ in range(all_tasks.options.reconstructor_options.num_epochs):
    for i_angle, angle in enumerate(angles):
        # Get the projection of the 3D object at the current angle.
        orig_object_proj = get_projection(object_3d, angle)

        # Overwrite the object in the current task manager with the projection.
        task.reconstructor.parameter_group.object.set_data(orig_object_proj)

        # Run one epoch with the current task manager
        task.run(1)

        # Take out updated 2D object.
        updated_object_proj = task.get_data("object")

        # Calculate the delta of the projection, backproject it, then
        # apply it to the 3D object.
        delta_object_proj = updated_object_proj - orig_object_proj
        delta_object_3d = backproject([n_x, n_y, n_z], angle, delta_object_proj)
        object_3d = object_3d + delta_object_3d
\end{Verbatim}

The last example demonstrates the usage of Pty-Chi in a plug-and-play framework that combines model-based phase retrieval and generative image editing for artifact suppression \cite{Du2025-co}. Plug-and-play is a variant of alternating direction method of multipliers (ADMM) \cite{Boyd2010-ia, Venkatakrishnan2013-ak} that splits a complex problem into several subproblems. In each epoch of plug-and-play, the solutions of all problems are updated sequentially, and the dual variables that coordinate the solutions of the subproblems are updated at the end. This process is repeated until the solutions of all subproblems reach a consensus. In the mentioned work, the update of the solution of the phase retrieval subproblem is performed with the following equation:
\begin{equation}
    o \leftarrow \argmin_o \left\{ f(o, \Theta) + \frac{\admmpenalty}{2}\norm{o - v + u}^2 \right\}
    \label{eq:pnp}
\end{equation}
where $f$ is the objective function of the phase retrieval subproblem, $o$ is the object function of the phase retrieval subproblem, $\Theta$ is the set of other reconstruction parameters (\emph{e.g.}, probe, probe positions), $v$ is the solution of the image editing subproblem, $u$ is the dual variable, and $\admmpenalty$ is the ADMM penalty factor. Rather than re-deriving the update rules to solve Eq.~\ref{eq:pnp} exactly, the cited work runs several epochs of Pty-Chi to update the object and other reconstruction parameters; the minimization of the norm in Eq.~\ref{eq:pnp} is realized by modifying the object in Pty-Chi's task manager after each phase retrieval epoch. The following pseudocode illustrates the code implementation:
\begin{Verbatim}[frame=single]
for _ in range(n_inner_epochs):
    task.run(1)
    o = task.get_data("object")
    o = o - tau * (o - v + u)
    task.reconstructor.parameter_group.object.set_data(o)
\end{Verbatim}

This example again shows how Pty-Chi's object-oriented design allows it to be directly used as the building block for other algorithms without any internal modification.

\section{Examples}

\subsection{Case study 1: compact batching and momentum acceleration in LSQML}

\begin{figure}
    \centering
    \includegraphics[width=1\linewidth]{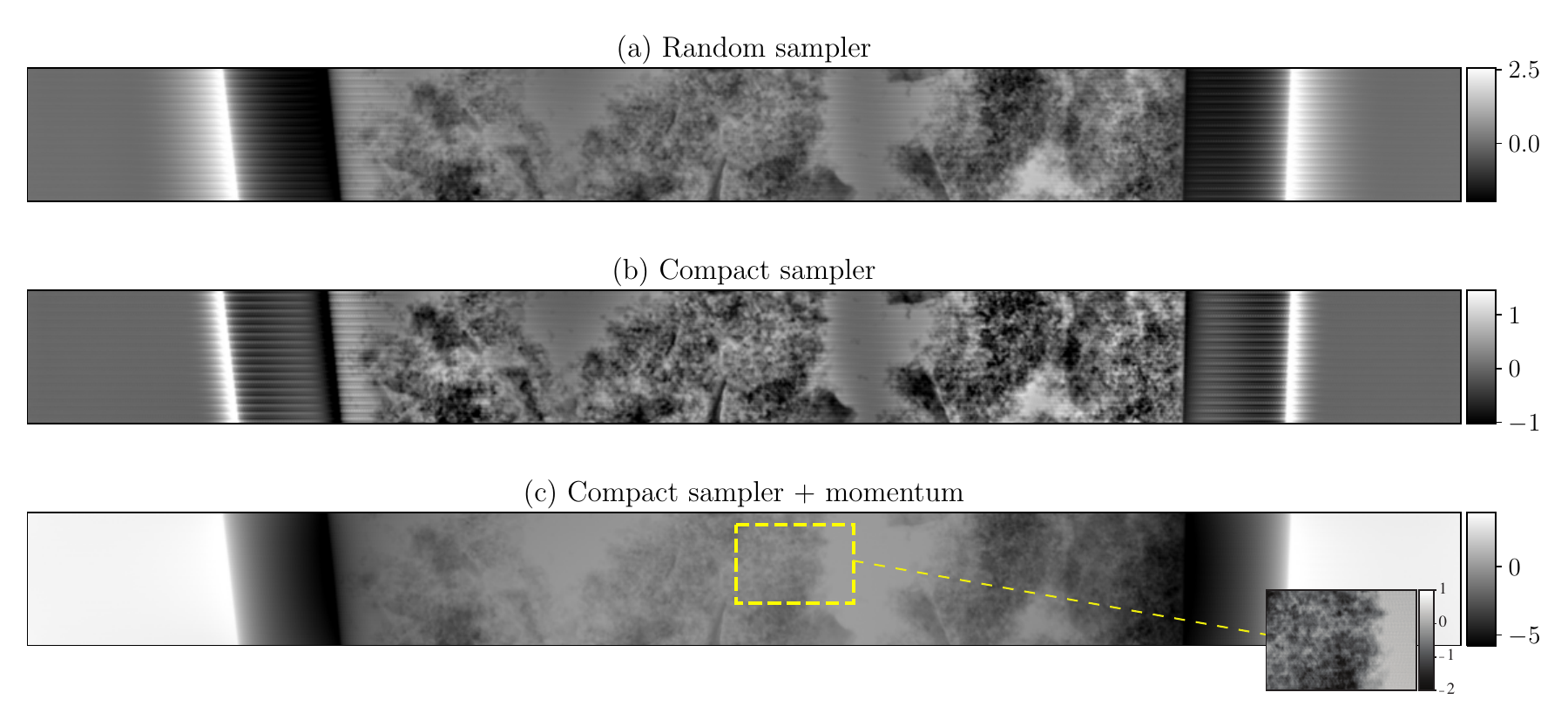}
    \caption{Reconstructions delivered by the LSQML reconstructor of Pty-Chi with various settings. (a) Random batch sampler. (b) Compact batch sampler. (c) Compact batch sampler with momentum acceleration. The former two cases exhibit halo around the object and horizontal stripe artifacts around the wall of the capillary tube. A clean and artifact-free reconstruction was only obtainable with the configurations in (c). Due to the large dynamic range of the reconstructed phase in (c), features inside the pipette are presented with seemingly reduced contrast. We show a selected portion of the features with smaller dynamic range in the inset to show that the real contrast is preserved.}
    \label{fig:fly110_recons}
\end{figure}

We present the reconstruction results of Pty-Chi on a challenging ptychographic dataset to demonstrate the implementation of the LSQML algorithm, the compact batch sampler, and the LSQML-specific momentum acceleration \cite{Wakonig2020-ap}. The sample imaged is a section of the tip of a borosilicate glass pipette with an outer diameter of about 80 \micron. The wall thickness of the pipette at this portion is about 8 \micron. \ce{TiO2} and \ce{SiO2} micro-particles are deposited on the inner wall of the pipette. 
Ptychographic data were collected on the Velociprobe \cite{Deng2019-wh} at beamline 2-ID-D of the Advanced Photon Source (APS). The probe was formed using a zone plate with a beam energy of 10 keV. 9926 diffraction patterns were collected in fly-scan mode \cite{Deng2015-hq, Huang2015-fo} with the scan path following a snake-pattern. For reconstruction, the probe was cropped to $128\times 128$, giving a real-space pixel size of 30.2 nm.

Due to the fly-scan setup, the probe is only partially coherent, and the vertical spacing of the scan points is about 3 times the horizontal spacing. Additionally, the thick side wall of the pipette has a strong phase shift, which results in a large phase difference between the wall and the air outside the sample. All these factors pose challenges to ptychographic reconstruction. In all the experiments introduced below, we use 15 probe modes to model the partial coherence due to fly-scan. With the LSQML reconstructor and the random batch sampler, the reconstructed image exhibits a strong halo (spurious high intensity outside the sample's boundary) and artifacts in the form of horizontal stripes around the wall of the pipette, as shown in Fig.~\ref{fig:fly110_recons}(a). Using the compact batch sampler slightly reduces the range of the halo but does not help with the stripe artifacts [Fig.~\ref{fig:fly110_recons}(b)]. However, enabling momentum acceleration drastically improved the image quality, where the halo and stripes are effectively suppressed, as shown in Fig.~\ref{fig:fly110_recons}(c). The phase of the side walls of the pipette also became more realistic; the phase gradually becomes less negative when one moves towards the outside of the pipette wall as the projected wall thickness reduces. 

\subsection{Case study 2: orthogonal probe relaxation}
\label{sec:siemens_star}

\begin{figure}
    \centering
    \includegraphics[width=0.8\linewidth]{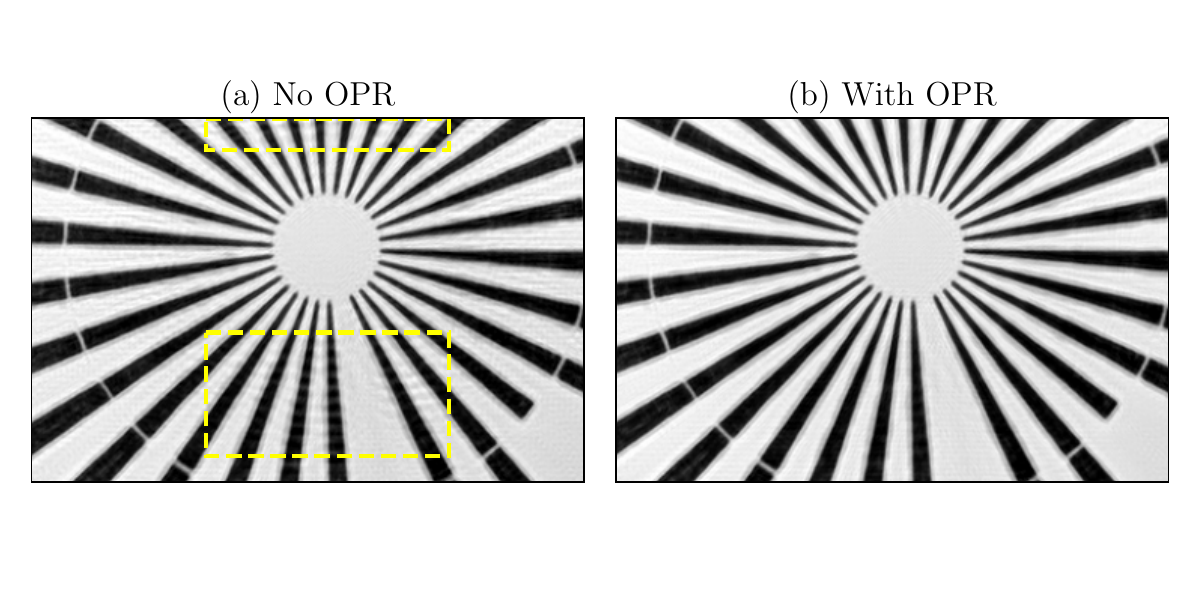}
    \caption{Reconstructions delivered by LSQML with (a) and without (b) OPR. The reconstruction without OPR exhibit ripple-like artifacts [highlighted by dashed boxes in (a)] due to the variation of illumination across scan lines. By modeling this variation with OPR, the artifacts are effectively suppressed.}
    \label{fig:fly85_recons}
\end{figure}

In this example, we validate the orthogonal probe relaxation (OPR) \cite{Odstrcil2016-wy} in Pty-Chi. The data used for this case were collected at APS 2-ID-D, and the sample is a Siemens star test object. The dataset for reconstruction involves 6732 diffraction patterns cropped to $128\times 128$. Reconstruction results introduced below were obtained with 30 incoherent probe modes. 

During the experiment, the imaging system experienced beam fluctuation, which caused the probe function to vary spatially. With the LSQML reconstructor, reconstructing the dataset without OPR modeling resulted in the object shown in Fig.~\ref{fig:fly85_recons}(a), where vertically varying ripple-like artifacts are present as a result of changing illumination throughout the course of the scan. By enabling OPR with 9 coherent modes (including the primary mode), the spokes of the Siemens star become more uniform, and the ripple artifacts are significantly reduced [Fig.~\ref{fig:fly85_recons}(b)]. 

\subsection{Case study 3: deep image prior-represented object}

\begin{figure}
    \centering
    \includegraphics[width=0.8\linewidth]{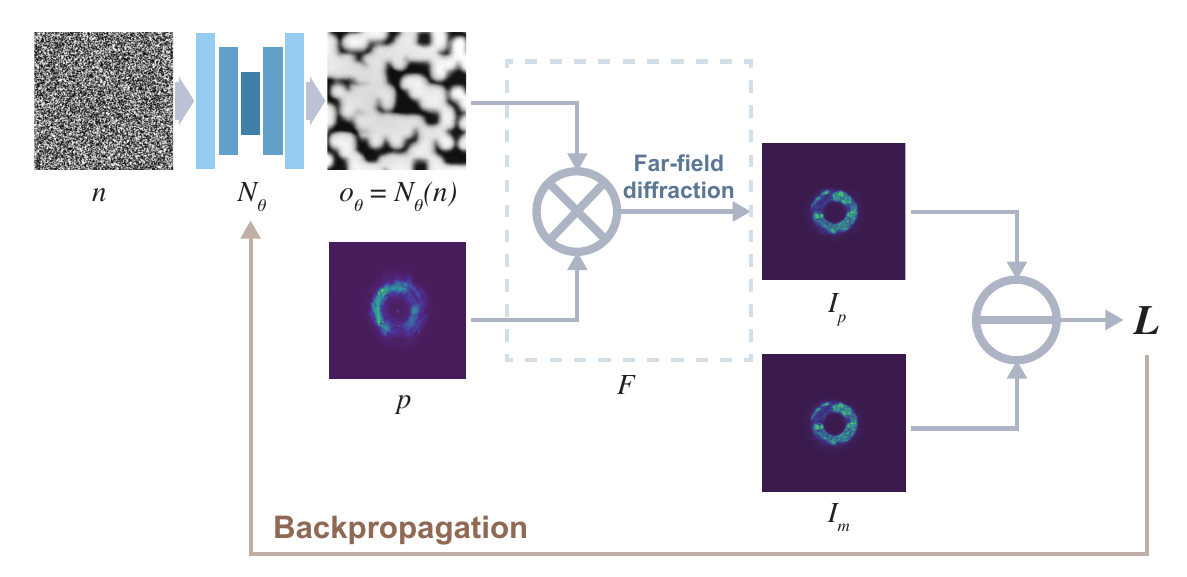}
    \caption{An illustration of reconstruction based on deep image prior. Instead of directly solving the object function, Pty-Chi's AD reconstructor solves for the parameters of a neural network that maps a fixed noise image into the supposed object. The rest of the forward model stays the same as traditional model-based reconstruction.}
    \label{fig:dip_diagrm}
\end{figure}

\begin{figure}
    \centering
    \includegraphics[width=1\linewidth]{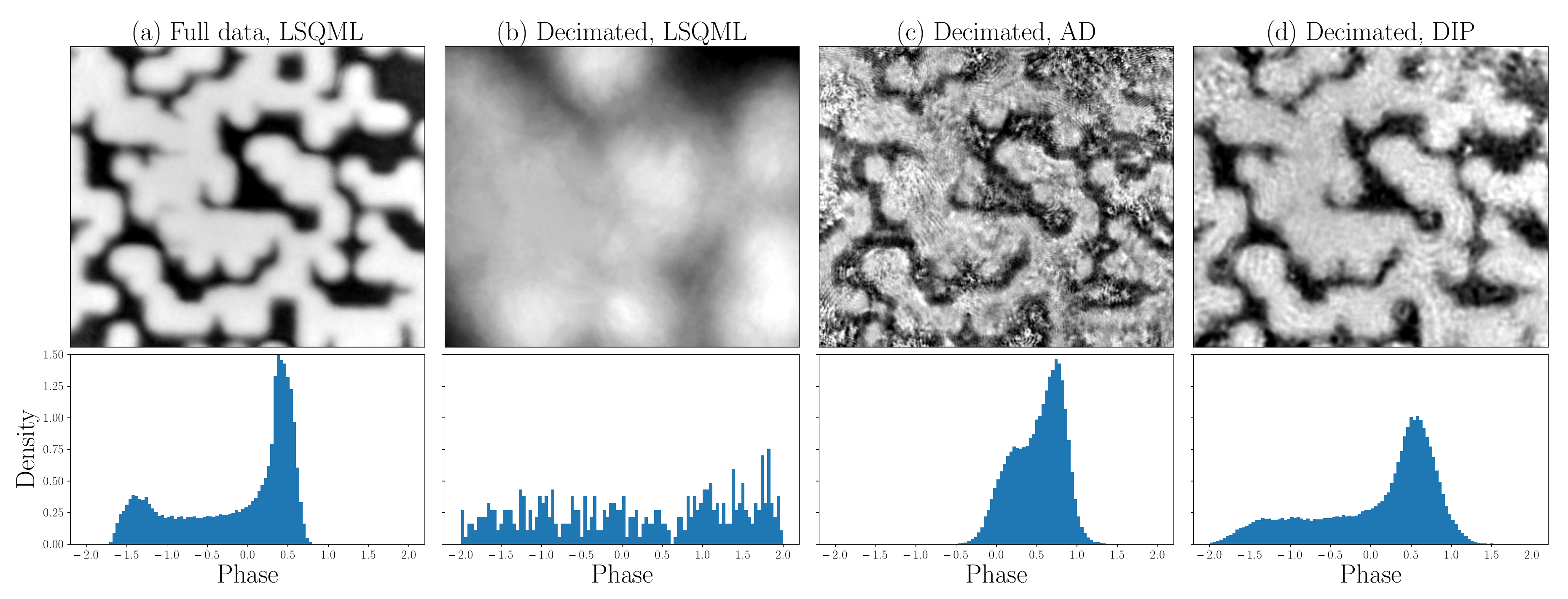}
    \caption{Reconstructed phase images of the full and decimated (keeping only 2\% of the original diffraction patterns) data. (a) The reconstruction of the full dataset obtained with LSQML, served as the reference image. (b-d) shows the reconstructions of the decimated dataset with LSQML, native AD, and AD with DIP-represented object. The histograms of the phase in the range of [-2, 2] are shown under the images.}
    \label{fig:dip_recons}
\end{figure}

We demonstrate an extended feature of Pty-Chi where the object being reconstructed is parameterized by a neural network instead of represented directly as a pixelated array, which allows one to enhance the reconstruction quality under adverse conditions by exploiting the deep image prior (DIP) characteristic of neural networks \cite{Lempitsky2018-ee}. Fig.~\ref{fig:dip_diagrm} shows the workflow of DIP-based reconstruction. An image-to-image neural network $N_\theta$ with parameters $\theta$ is created and randomly initialized. A Gaussian random image $n$ is also generated at the beginning of the reconstruction and kept constant. The neural network maps $n$ into the supposed object function, $o_\theta = N_\theta(n)$. The rest of the forward model is identical to the traditional ptychographic image formation model, where the generated object $o_\theta$ is used in the same way as the one using a pixelated representation. The predicted intensity is thus
\begin{equation}
    I_p = F\left(N_\theta(n), p, \Lambda\right)
\end{equation}
where $F(o, p, \Lambda)$ is the ptychographic forward model taking the object, probe, and other parameters ($\Lambda$) as arguments. The loss function $L$ is evaluated using $I_p$ and measured intensity $I_m$, and the gradient $\partial L / \partial \theta$ is calculated and used to update $\theta$, tuning the network parameters so that it generates an object that better conforms with the data and physics. 

In Pty-Chi, we build deep image prior-based reconstruction on the AD reconstructor, directly reusing most of its components except for the input object to the forward model, which is replaced with the $o_\theta$ generated by $N_\theta(n)$. The AD engine of PyTorch automatically computes $\partial L / \partial \theta$. 

The tendency of neural networks to generate low-entropy, low-noise ``natural images'' makes the DIP method particularly useful in reconstructing poorly sampled data. We demonstrate this advantage using a dataset collected at the Hard X-ray Nanoprobe (26-ID) of the APS, where the sample is a randomly etched test pattern. The specifications of the sample and the data collection experiment have been documented in \cite{Babu2023-pq} and \cite{Du2024-qp}. The dataset originally contains 961 scan points with an average nearest-neighbor distance of 55 nm (calculated from the actual positions). We decimated the data by 50 times (keeping just one of every 50 scan points), leaving only 2\% of the original data or 20 diffraction patterns; the new average nearest-neighbor distance is 387 nm. Considering that the diameter of the probe is 440 nm, this reduces the overlap ratio to merely 12\%. We reconstructed the full data using LSQML, and the 50x decimated data using LSQML, AD, and DIP-enabled AD. For DIP, the neural network used to represent the object is a 4-level ``hourglass'' (U-Net \cite{Ronneberger2015-do} without skip connections) which takes a Gaussian noise image with 32 channels and the same lateral size as the object as input, and outputs a monochannel image with the same lateral size for both phase and magnitude. 

We show the reconstruction obtained with full and decimated data in Fig.~\ref{fig:dip_recons}. The reconstructed image with the full data [Fig.~\ref{fig:dip_recons}(a)] reveals that the sample is nearly binary with clearly defined boundaries between both phases. After decimation, LSQML (b) failed to deliver any sensible structure. Using the Adam optimizer \cite{Kingma2014-dm}, AD (c) recovered the overall contours of the features, but the image is heavily degraded by noise and spurious inhomogeneity in regions that are supposed to be uniform. On the other hand, DIP-enabled AD significantly improved the contrast and suppressed the high-frequency artifacts as shown in (d). While there are still artifacts due to the lack of overlap constraint, the phase histogram of (d) exhibits a bimodal distribution more similar to (a) and with a closer contrast. 

\section{Performance benchmarking}

In this section, we present the performance benchmarking results of Pty-Chi. All tests shown in this section were conducted on a Supermicro SYS-740GP-TNRT server with two 28-core 2.60 GHz Intel Xeon Gold 6348 CPUs and four NVIDIA H100 GPUs. Additional benchmarking results on the speed of Pty-Chi with different GPUs can be found in the supplemental document. 

\begin{figure}
    \centering
    \includegraphics[width=0.9\linewidth]{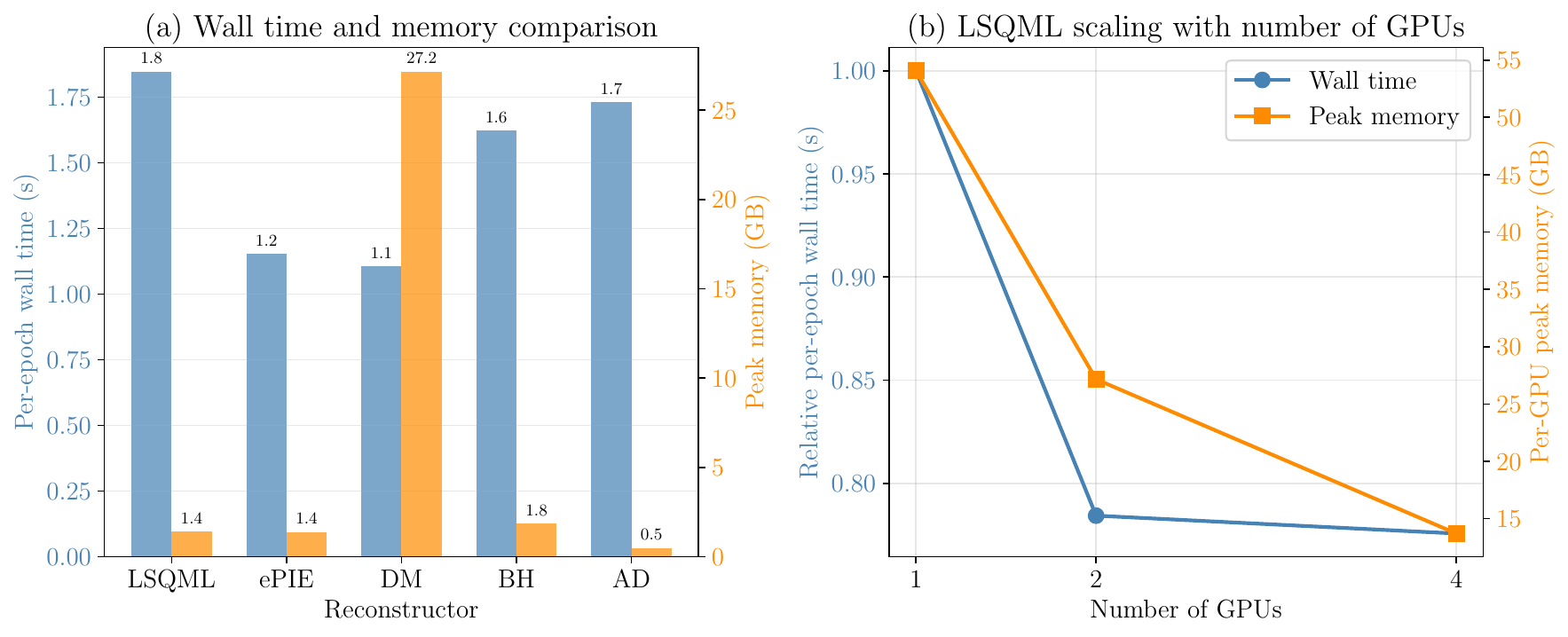}
    \caption{Performance benchmarking results of Pty-Chi. (a) The per-epoch wall time and peak GPU memory during 100 epochs of reconstructions on the dataset shown in Section \ref{sec:siemens_star}. The dataset contains 6732 diffraction patterns, each of size $128\times 128$. The object size was set to $413\times 535$. 30 probe modes and a batch size of 50 were used. Since not all reconstructors support OPR, OPR was not used in these tests. (b) The relative wall time as a fraction of the single-GPU case and peak GPU memory (per-GPU) of LSQML using 1, 2, and 4 GPUs, tested on the same dataset. Different from (a), 9 OPR modes (including the primary mode) were used in these tests, and the batch size was set to 2000 for more speedup and less overhead.}
    \label{fig:benchmark}
\end{figure}

We first compare the reconstruction speed and peak GPU memory usage across the reconstructors available in Pty-Chi. The tests were performed on the dataset shown in Section \ref{sec:siemens_star}, which contains 6732 diffraction patterns. The size of each diffraction pattern is $128\times 128$. Based on the scan range and the pixel size, an object size of $413\times 535$ was used. Since not all reconstructors support OPR, OPR was not used in these tests. The reconstructions were run with a single H100 GPU. Fig.~\ref{fig:benchmark}(a) shows the average reconstruction wall time per epoch and peak GPU memory over 100 epochs of LSQML, ePIE (representing the PIE family), DM, BH, and AD reconstructors. In terms of wall time, LSQML is relatively more expensive than other algorithms because the construction of the object preconditioner \cite{Odstrcil2018-ns} incurs extra cost. ePIE and DM have the lowest per-epoch wall time. However, because it stores more intermediate variables, DM requires significantly more memory. We find that AD is the least memory-demanding reconstructor. This may seem counterintuitive given the common myth that AD consumes more memory because it must save the output of each non-linear operator during the forward pass. However, such storage is also required by analytical reconstructors, and modern AD engines like PyTorch employ multiple strategies such as reference counting \cite{Paszke2019-xm} to reduce memory usage.

In Fig.~\ref{fig:benchmark}(b), we test the wall time (measured as fractions of the single-GPU case) and memory scaling performance with multiple GPUs. The tested reconstructor was chosen to be LSQML. We used the same dataset as was used in (a), but 9 OPR modes, including the primary mode, were used. Also, the batch size was increased to 2000 to reduce synchronization frequency and communication overhead. Reconstruction was run with 1, 2 and 4 GPUs with the total batch size held constant, so each GPU received fewer diffraction patterns as the GPU count increased (\emph{i.e.}, a strong scaling test). The per-epoch wall time decreased by 22\% with 2 GPUs. However, the reduction plateaued at 4 GPUs because communication overhead became significant relative to the speedup. On the other hand, the per-GPU peak memory usage is nearly inversely proportional to the number of GPUs, signifying the direct effect of multi-GPU parallelization on memory reduction.

\section{Conclusion}

We have developed Pty-Chi, a modern ptychographic reconstruction package. Through the use of PyTorch's inherent capabilities, rigorously implemented reconstruction algorithms, and object-oriented architecture design, Pty-Chi provides both high-performance analytical reconstructors and flexible automatic differentiation reconstructors, supports GPU acceleration and multi-device parallelization, offers room for capability innovation through creative usage, and can serve as the building block for the development of novel algorithms and workflows. We are aware of the adoption of Pty-Chi at multiple synchrotron radiation facilities and universities at the time this report is written. With an open-source and open-contribution model, we are expecting further community-supported development of Pty-Chi as well as more innovative methods and techniques derived from and built upon the software. 

\section*{Code availability}

The Pty-Chi project is located at \url{https://github.com/AdvancedPhotonSource/Pty-Chi}.

\section*{Acknowledgments}

Work performed at the Advanced Photon Source and Center for Nanoscale Materials, both U.S.~Department of Energy Office of Science User Facilities, was supported by the U.S.~DOE, Office of Basic Energy Sciences, under Contract No.~DE-AC02-06CH11357.

\printbibliography

\end{document}


\maketitle

\section{Speed benchmarking with various GPUs}
\begin{figure}[h]
    \centering
    \includegraphics[width=1\linewidth]{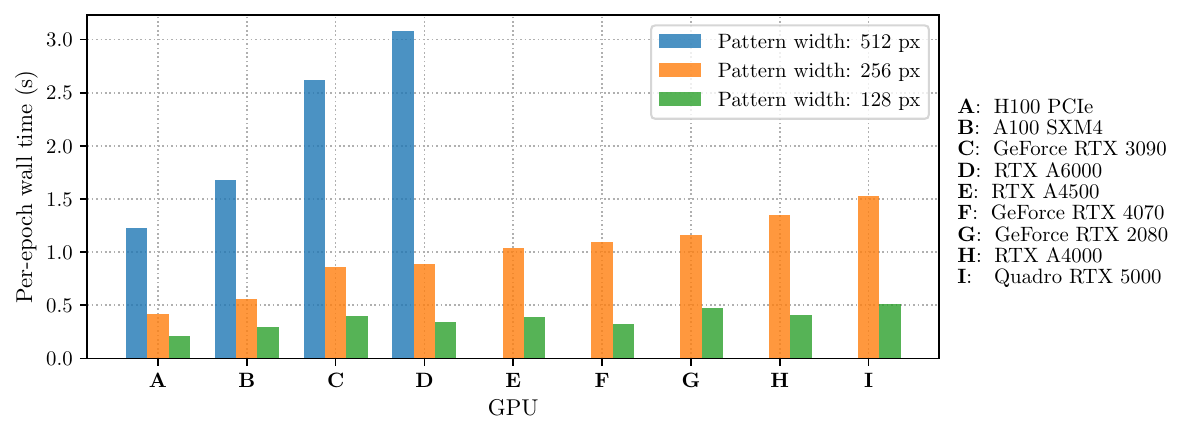}
    \caption{The per-epoch wall time of Pty-Chi's LSQML reconstructor with various GPUs and diffraction pattern sizes. Only one GPU was used for each test. }
    \label{fig:speed_various_gpus}
\end{figure}

The per-epoch wall time was measured on 9 different GPUs, as shown in Fig.~\ref{fig:speed_various_gpus}. The benchmarking tests were run with the following properties and settings:
\begin{itemize}
    \item Reconstructor: LSQML
    \item Number of diffraction patterns: 961
    \item Batch size: 16
    \item Probe size: 128 $\times$ 128
    \item Number of probe modes: 15
    \item Number of OPR modes: 3, including the primary mode
    \item Position correction: disabled
\end{itemize}
GPUs E to I do not have measurements for the 512 $\times$ 512 diffraction pattern size because the memory needed exceeds what is available on those GPUs. 

The host-device transfer speeds vary from system to system, but the transfer time is negligible for the purposes of this comparison. The execution time of memory operations (host-to-device transfer, device-to-host transfer, device-to-device transfer, and memset) was measured for the A100 SXM4 and RTX A4000 GPU and was in the range of 3-6\% of the total execution time, with the host-to-device memory transfer taking up most of the memory operation time.

